\documentclass[twocolumn,twoside]{article}
\usepackage{graphics}
\newcommand{\hb}{\\ \hspace*{2ex}}

\begin{document}
\title{DETERMINATION OF THE GALAXY CLUSTER ORIENTATION USING X-RAY IMAGES BY FOCAS METHOD}
\author{S.Yu.\,Shevchenko$^{1}$, A.V.\,Tugay$^{2}$\\[2mm] %English only
\begin{tabular}{l}
 $^{1}$ Schmalhausen Institute of Zoology, \hb
 Bohdan Hmelnitskiy St., 15, Kyiv, Ukraine, {\em astromott@gmail.com}\\
 $^{2}$ Astronomy and Space Physics Department, Faculty of Physics, \hb 
 Taras Shevchenko National University of Kyiv,\hb
 Glushkova ave., 4, Kyiv, 03127, Ukraine,  {\em tugay.anatoliy@gmail.com}\\
\end{tabular}
}
\date{}
\maketitle

ABSTRACT.

In our work we considered orientations of bright X-ray halos of the  galaxy clusters (mainly Abell clusters). 
78 appropriate clusters were selected using data from Xgal sample of extragalactic objects in XMM-Newton observation archive. Position angles and eccentricities of these halos were calculated applying FOCAS method. No privileged orientations were found.

{\bf Keywords}: Galaxies: clusters; X-rays: galaxies: clusters.
\\[3mm]

{\bf 1. Introduction}\\[1mm]

One of the interesting tasks of the extragalactic astronomy is a search of “dedicated” directions in the galaxies and their clusters orientations.
The catalog of Abell Cluster Objects (Abell et al., 1989) is the main catalog of galaxy clusters. It contains the most of closest and brightest clusters which are the most suitable for both optical and X-ray observations.
Galaxy orientations in 247 rich Abell clusters were studied by Godlowski et al. (2010) and Panko et al. (2013) with corresponding statistical data analysis and simulations. 
Orientation of the galaxies from relatively small sample can be numerically described by the distribution of anisotropy parameter. 
This parameter was calculated for edge-on galaxies in Parnovsky \& Tugay (2007) and for nearby galaxy groups in Godlowski et al.(2012). 
Galaxies orientation in the nearby groups was studied by Pajowska et al.(2012).

X-ray images of galaxy cluster halos could be easily approximated by ellipses, so they are well suitable to study such large-scale orientation.
We use XMM-Newton archive based sample of X-ray extragalactic objects X-Gal (Tugay, 2012). 
In the previous work (Tugay et al., 2016) the orientation of 30 X-ray southern clusters from PF catalog (Panko \& Flin, 2006) was determined using the method of image isophotes approximation by ellipses. 
In this work we estimated orientation of the whole sky sample of X-ray brightest galactic clusters with more accurate FOCAS method (Jarvis\& Tyson, 1981).\\ [2mm]

%\pagebreak

{\bf 2. Method}\\[1mm]

Primarily X-Gal sample, comprised by about 5 000 objects was examined to find the brightest galaxy clusters. 
As a result 77 clusters were selected applying the precondition of the most right in X-Ray band. 
We used $1x10^{-11}$ - $1х10^{-12}$ $mW/m^2$ interval for the flux in 2-10 keV range.
The major part of these objects (70$\% $) – turned out to be Abell clusters. Redshifts of our clusters are in the range from z=0,02 to 0,3 (i.e. their distances reaches up to 1 Gpcs).
XMM-Newton imaging data (in FITS format) were retrieved from LEDAS (LEicester Database and Archive Service) and processed by XMM-SAS software.
The next step was to use these data to calculate clusters’ X-Ray halo position angles (PA), eccentricities and their inaccuracies using FOCAS algorithm.
Results are shown in Tables 1 and 2. The clusters in the tables are arranged by right ascension.
For the obtained results we determined equatorial coordinates of the clusters orientation effective vector $\vec{n_A}$ (Fig. 1).  Vector $\vec{n}$ points to the cluster center. Vector $n_z$ points to Northen celestial pole – (0, 0, 1).
Cluster ellipsoid projection is given on the Fig. 2, where $\vec{n_A}$ vector is the direction of the main axis of visible image.
Direction of the third ellipsoid axis $n_3$ is not known. But it is known that the cluster ellipticity is insignificant and if there is measurable eccentricity of the cluster then it would be in the range of existing axis $n_A$ and $n_N$. 
In general case, spatial orientation of the vector $\vec{n_3}$  is perpendicular to $\vec{n_A}$ and hence within single-valued conversion of $\vec{n_3}$ it is possible to use vector $\vec{n_A}$ instead. 
Hence vector $\vec{n_A}$ was used to describe possible orientations of the cluster. 
It was determined from the following equations:

\begin{equation}
 \left\{ 
 \begin{array}{r}
  \vec{n_N} \cdot [\vec{n} \times \vec{n_Z} ]=0 \\
  \vec{n_N} \times \vec{n}=0  \\
 \end{array}
\right.
\end{equation}

\begin{equation}
 \left\{ 
 \begin{array}{l}
  \vec{n_A} \times \vec{n}=0 \\
  \vec{n_A} \times \vec{n_N}=cos(PA)  \\
 \end{array}
\right.
\end{equation}  \hb 
 \begin{figure}[h]
 \resizebox{\hsize}{!}{\includegraphics{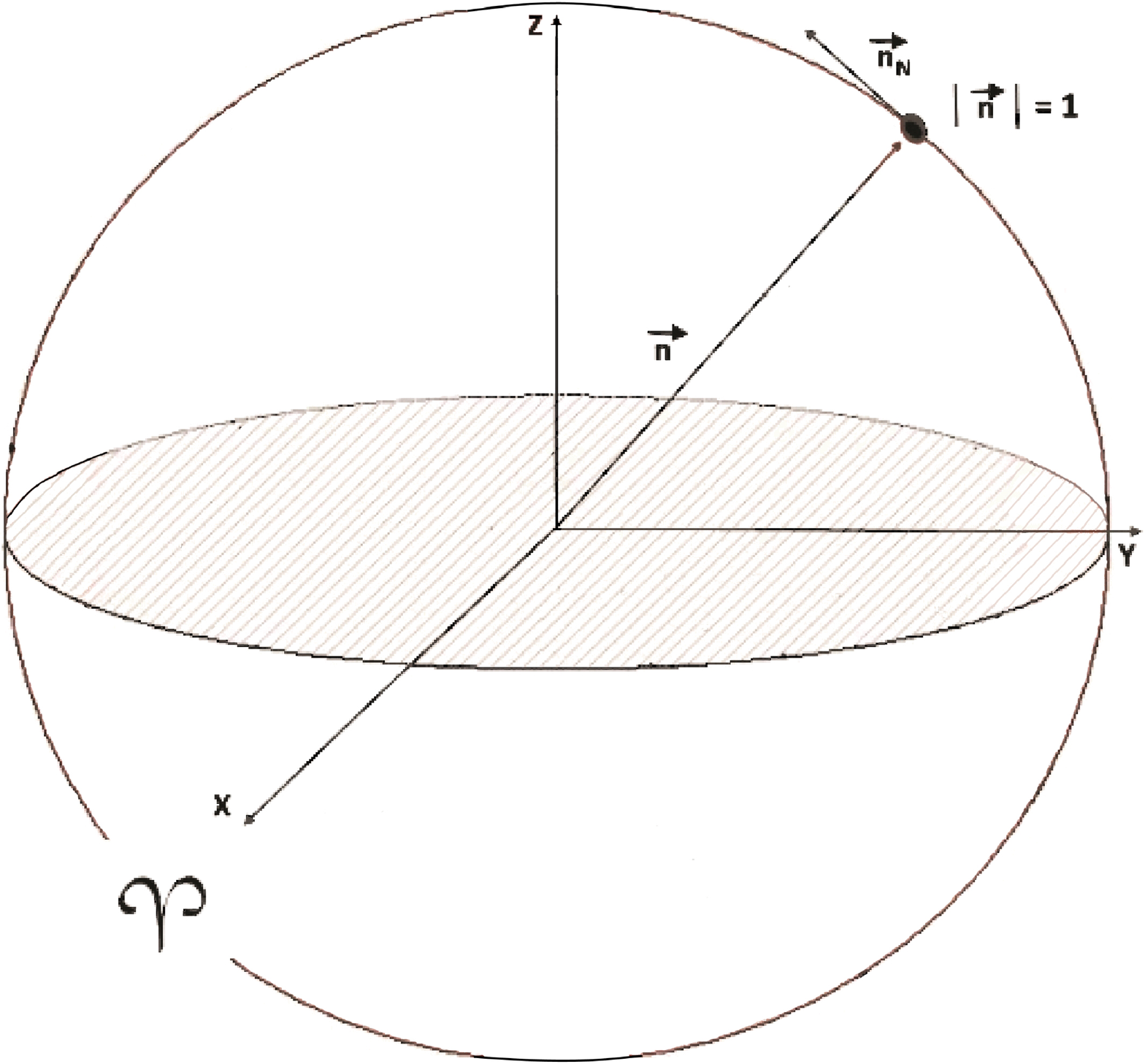}}
 \caption{Cluster orientation directions.}
 \end{figure}

 \begin{figure}[h]
 \resizebox{\hsize}{!}{\includegraphics{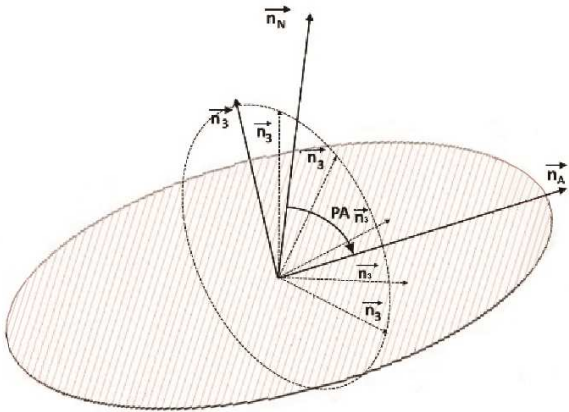}}
 \caption{X-Ray halos orientation directions plane projections.}
 \end{figure}

\begin{figure}[h]
 \resizebox{\hsize}{!}{\includegraphics{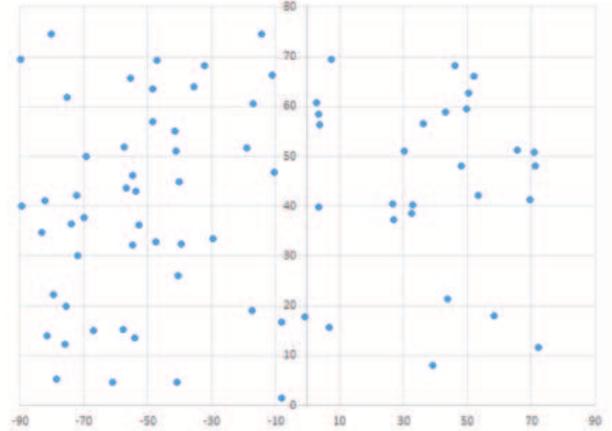}}
 \caption{Distribution of the X-Ray halos orientation directions for RA and DEC.}
 \end{figure}

\begin{table}[h]
\caption{Orientations and Eccentricities of X-Ray halos.}
\begin{tabular}{lrr}
\hline
 Name                           &    PA              &  e                 \\
\hline
ACO 2700                        &    25  $\pm $   11  &  0.51 $\pm $  0.2\\         
ACO 119                         &   110  $\pm $    3  &  0.09 $\pm $ 0.01\\         
ACO 122                         &    84  $\pm $    2  & 0.713 $\pm $0.327\\         
ACO 2984                        &    84  $\pm $    1  &  0.52 $\pm $  0.4\\         
ACO 399                         &    63  $\pm $   15  &  0.31 $\pm $  0.1\\         
ACO 401                         &    53  $\pm $    1  &  0.39 $\pm $ 0.25\\         
ACO 3112                        &    71  $\pm $    5  &  0.62 $\pm $  0.5\\         
ACO 3158                        &    82  $\pm $    4  & 0.212 $\pm $ 0.01\\         
ACO S 384                       &    38  $\pm $   20  &  0.36 $\pm $ 0.25\\         
ClG 0422-09                     &    19  $\pm $    1  & 0.079 $\pm $ 0.01\\         
\hline
ACO 496                         &     8  $\pm $    3  &   0.2 $\pm $  0.1\\         
ClG 0451-03                     &    71  $\pm $    8  &  0.17 $\pm $ 0.02\\         
MCXC J0528.9-3927               &    46  $\pm $   39  &  0.28 $\pm $ 0.16\\         
MCXC J0532.9-3701               &    88  $\pm $    1  &  0.19 $\pm $ 0.14\\         
ACO 3378                        &   111  $\pm $   20  &  0.45 $\pm $ 0.21\\         
ACO 3391                        &   107  $\pm $    3  &  0.22 $\pm $ 0.04\\         
ACO 3404                        &    45  $\pm $    2  &  0.47 $\pm $  0.3\\         
ZwCl 0735+7421                  &    42  $\pm $   11  &  0.15 $\pm $ 0.08\\         
ClG 0745-1910                   &    76  $\pm $    2  &  0.17 $\pm $ 0.46\\         
ACO 653                         &    46  $\pm $    3  &  0.36 $\pm $ 0.2 \\         
\hline
ACO 689                         &    46  $\pm $   34  &  0.14 $\pm $ 0.01\\         
ACO 773                         &   103  $\pm $    5  &  0.37 $\pm $ 0.25\\         
ACO 901A                        &    41  $\pm $   20  &  0.36 $\pm $ 0.27\\         
ACO 907                         &    45  $\pm $   22  &  0.42 $\pm $  0.3\\         
ZwCl 1021+0426                  &    35  $\pm $    9  &  0.46 $\pm $  0.4\\         
ACO 1084                        &    20  $\pm $    6  &  0.51 $\pm $ 0.33\\         
ACO 1201                        &    25  $\pm $   15  &   0.5 $\pm $  0.2\\         
ClG J1115+5319                  &    66  $\pm $     5  & 0.25 $\pm $ 0.02 \\         
ACO 1413                        &     4  $\pm $    2  &  0.63 $\pm $ 0.35\\         
\hline
MCXC J1206.2-0848               &    59  $\pm $   25  &  0.24 $\pm $ 0.17\\         
ZwCl 1215+0400                  &   144  $\pm $    9  &  0.49 $\pm $ 0.15\\         
ACO S 700                       &     8  $\pm $    3  &  0.07 $\pm $ 0.01\\         
ACO 3528                        &    10  $\pm $    1  &   0.6 $\pm $  0.4\\         
ACO 1651                        &    99  $\pm $    3  &  0.35 $\pm $ 0.27\\         
ACO 1656                        &    73  $\pm $    3  &  0.18 $\pm $  0.1\\         
ACO 1663                        &    48  $\pm $   39  &   0.3 $\pm $  0.2\\         
ACO 1664                        &   157  $\pm $    5  &  0.48 $\pm $ 0.25\\         
2E 2975                         &    60  $\pm $   15  &  0.28 $\pm $ 0.25\\         
1325-5737 &    45  $\pm $   33  &  0.17 $\pm $  0.1\\         
\hline
ACO 3558                        &    43  $\pm $   16  &  0.47 $\pm $  0.3\\         
ACO 1750N                       &   140  $\pm $   25  & 0.379 $\pm $ 0.24\\         
ACO 3560                        &    26  $\pm $   14  &  0.32 $\pm $ 0.27\\         
ACO 3562                        &   150  $\pm $   29  &  0.31 $\pm $ 0.14\\         
ACO 1775                        &   136  $\pm $   40  &  0.31 $\pm $ 0.23\\         
ACO 3571                        &    84  $\pm $    2  &   0.3 $\pm $ 0.01\\         
ClG J1347-1145                  &    11  $\pm $    6  &  0.15 $\pm $ 0.08\\         
ACO 1835                        &    10  $\pm $    5  &  0.21 $\pm $ 0.16\\         
ACO 3581                        &    69  $\pm $    7  &   0.2 $\pm $ 0.06\\         
1419+2511    &    30  $\pm $    3  &  0.45 $\pm $  0.3\\         
\hline
NGC 5718 Group                  &    61  $\pm $    2  &  0.22 $\pm $ 0.15\\         
ClG J1504-0248                  &   147  $\pm $   17  &   0.1 $\pm $ 0.06\\         
ACO 2050                        &   141  $\pm $   34  &  0.47 $\pm $ 0.25\\         
ACO 2052                        &   135  $\pm $    6  &  0.48 $\pm $  0.3\\         
ACO 2051                        &    64  $\pm $    8  &  0.25 $\pm $ 0.03\\         
ACO 2055                        &    12  $\pm $   12  &  0.31 $\pm $ 0.13\\         
ACO 2063                        &     8  $\pm $    4  &  0.25 $\pm $ 0.06\\         
ACO 2204                        &     7  $\pm $    5  & 0.087 $\pm $ 0.01\\         
ClG J1720+2638                  &    70  $\pm $    3  &  0.38 $\pm $ 0.01\\         
MCXC J2011.3-5725               &    60  $\pm $   15  &  0.28 $\pm $ 0.25\\         
\hline
\end{tabular}
\end{table}

\begin{table}[h]
\caption{The same as Table 1 for the last clusters.}
\begin{tabular}{lrr}
\hline
 Name                           &    PA              &  e                 \\
\hline
ACO 3667                        &     8  $\pm $    8  &  0.76 $\pm $ 0.08\\         
2MAXI J2014-244                 &     9  $\pm $    1  &  0.52 $\pm $ 0.45\\         
ACO 3693                        &    39  $\pm $   23  &  0.12 $\pm $ 0.08\\         
ClG J2129+0005                  &    24  $\pm $    1  &   0.5 $\pm $  0.3\\         
ACO 3814                        &    65  $\pm $   23  &  0.56 $\pm $ 0.27\\         
2217-1725 & 71$\pm $ 6&0.22$\pm $0.2\\         
ACO 3854                        &    37  $\pm $   22  &  0.51 $\pm $ 0.37\\         
ACO 3856                        &    39  $\pm $   10  &  0.56 $\pm $ 0.31\\         
ACO S 1101                      &   121  $\pm $    2  &  0.45 $\pm $  0.3\\         
ACO 3992                        &     2  $\pm $    2  &  0.37 $\pm $  0.1\\         
\hline
ACO 2597                        &    37  $\pm $    1  &  0.17 $\pm $ 0.05\\         
ACO 4010                        &   148  $\pm $    4  & 0.148 $\pm $ 0.01\\         
ACO 2626                        &   156  $\pm $    5  &  0.67 $\pm $ 0.06\\         
ACO 2667                        &   133  $\pm $   27  & 0.403 $\pm $ 0.28\\         
ACO 2670                        &   132  $\pm $   35  &  0.23 $\pm $ 0.14\\         
ACO 13                          &    60  $\pm $   20  &  0.41 $\pm $  0.1\\         
0018-0053                       &   149  $\pm $    2  &  0.12 $\pm $ 0.05\\         
ClG 0016+16                     &    67  $\pm $   24  &  0.18 $\pm $ 0.02\\               
\hline
\end{tabular}
\end{table}

\begin{table}[h]
\caption{Coordinate codes for clusters with long designations.}
\begin{tabular}{lr}
\hline
 Name                     & Coordinate code \\
\hline
1RXS J132441.9-573650     & 1325-5737 \\
2XMM J141830.6+251052     & 1419+2511 \\
XMMXCS J221656.6-172527.2 & 2217-1725 \\         
2XMMi J001737.3-005239    & 0018-0053 \\         
\hline
\end{tabular}
\end{table}

{\bf 3. Results and conclusion}\\[1mm]

Orientation and Eccentricities of X-Ray halos are presented in Tables 1-2 and the distribution of the $\vec n_A$ orientation (RA, Dec) is given on Fig.3. 
To check possible anisotropy the range of the right accession and declination of the $\vec{n_A}$ was split into 10 degree intervals. Then using Kolmogorov's criterion we determined maximal mean deviations for clusters number in each interval and hence further verified probability of the clusters orientation anisotropy.     
Uniformity of the distribution was established using above Kolmogorovs’ criterion. Application of this criterion enables with probability up to $95\% $ to accept hypothesis that Local Universe in the range up to about 1 Gpcs is isotropic.  
Further comparisons of the X-ray halos alignments and  galaxies clusters orientation in optical band might be required and useful in the next works. \\[2mm]

{\it Acknowledgements.} The authors are thankful to Elena Panko for discussions related to this work. This research retrieved data from NASA's Astrophysics Data System and LEDAS archive of XMM-Newton observations. \\[1mm]

{\bf References\\[1mm]}
\noindent\\
Abell G.O., Corwin H.G., Olowin, R.P.: 1989, {\it ApJS}, {\bf 70}, 1.\\
Godlowski W., Piwowarska P., Panko E., Flin P.: 2010, {\it ApJ}, {\bf 723}, 985.\\
Godlowski M., Panko E., Pajowska P., Flin P.: 2012, {\it JPhSt.}, {\bf 16}, 3901.\\
Jarvis J.F., Tyson J.A.: 1981, {\it AJ}, {\bf 86}, 476 \\
Pajowska P., Godlowski M., Panko E., Flin P.: 2012, {\it JPhSt.}, {\bf 16}, 4901.\\
Panko E., Piwowarska P., Godlowska J., Godlowski W., Flin P.: 2013, {\it Ap.}, {\bf 56}, 322.\\
Parnovsky S., Tugay A.: 2007, {\it JPhSt.}, {\bf 11}, 366.\\
Tugay A.: 2012, {\it Odessa Astron. Publ.}, {\bf 25}, 142. \\
Tugay A., Dylda S., Panko E.: 2016, {\it Odessa Astron. Publ.}, {\bf 29}, 34. \\

\vfill

\end{document}